\documentclass[12pt]{article}%\documentstyle[12pt]{article}
\input epsf.sty
\usepackage{mathrsfs}
\usepackage{amsmath}
\usepackage{mathrsfs}
\usepackage{amssymb}
\usepackage{color}
\usepackage{psfrag}
\usepackage{graphicx}
\usepackage{subfigure}
\usepackage{rotating}

\newcommand{\bea}{\begin{eqnarray}}
\newcommand{\eea}{\end{eqnarray}}
\newcommand{\be}{\begin{equation}}
\newcommand{\ee}{\end{equation}}
\newcommand{\vs}[1]{\vspace{#1 mm}}

\newcommand{\dsl}{\pa \kern-0.5em /}

\newcommand{\pa}{\partial}

\newcommand{\nn}{\nonumber\\}

\begin{document}
\topmargin 0pt
\oddsidemargin 0mm

\begin{flushright}

%USTC-ICTS-10-12\\

%hep-th/yymmnnn\\

\end{flushright}

\vspace{2mm}

\begin{center}

{\Large Wilson loops in $(p+1)$-dimensional Yang-Mills theories\\
using gravity/gauge theory correspondence}

\vs{10}

{Somdeb Chakraborty\footnote{E-mail: somdeb.chakraborty@saha.ac.in} and 
Shibaji Roy\footnote{E-mail: shibaji.roy@saha.ac.in}}

 \vspace{4mm}

{\em

 Saha Institute of Nuclear Physics,
 1/AF Bidhannagar, Calcutta-700 064, India\\}

\end{center}

\vs{10}

\begin{abstract}
We compute the expectation values of both the time-like and the light-like
Wilson loops in a strongly coupled plasma of $(p+1)$-dimensional Yang-Mills 
theories using gravity/gauge theory correspondence. From the time-like 
Wilson loop we obtain the velocity dependent quark-antiquark potential 
where the dipole is moving through the plasma with an 
arbitrary velocity $0<v<1$ and also obtain expressions for the screening 
lengths. When the velocity $v \to 1$, the Wilson loop becomes light-like and
we obtain the form of the jet quenching parameter in those strongly coupled
plasma. 
\end{abstract}

\newpage

\section{Introduction}

AdS/CFT correspondence \cite{Maldacena:1997re,Witten:1998qj,Gubser:1998bc}
and its generalizations \cite{Aharony:1999ti} help us to access the
non-perturbative regimes of SU($N$) gauge theories at large $N$ simply from
the low energy, weakly coupled string theory in certain backgrounds. 
Wilson loops are non-perturbative objects in gauge theories and the precise
prescription for the computation of its expectation values using AdS/CFT 
correspondence has been given in 
\cite{Maldacena:1998im,Rey:1998ik,Rey:1998bq,Brandhuber:1998bs}.
In strongly coupled gauge theories 
of interacting quark-gluon plasma, Wilson loops can be related to 
various measurable quantities in heavy ion experiments in RHIC or in LHC.
For example, the expectation value of a special time-like Wilson loop can be
related to the static quark-antiquark potential \cite{Wilson:1974sk} in 
a moving quark-gluon
plasma. On the other hand, the expectation value of a particular light-like
Wilson loop can be related, among other things, to the radiative energy loss
of a parton or the jet quenching parameter \cite{Kovner:2003zj}. 

The velocity dependent quark-antiquark potential of a dipole moving with an 
arbitrary
velocity through the hot quark-gluon plasma including the screening length 
\cite{Liu:2006nn,Caceres:2006ta,Chernicoff:2006hi,Avramis:2006em} as well 
as the jet quenching parameter  
\cite{Liu:2006ug,Liu:2006he}\footnote{Also see \cite{CasalderreySolana:2011us}
for a recent review.} have been calculated when the
plasma is described by $D=4$, ${\cal N} = 4$, SU($N$) Yang-Mills theory 
using AdS/CFT correspondence\footnote{Jet quenching parameter in various other 
theories have been obtained in \cite{Buchel:2006bv}. Also the drag force 
on a moving quark have been calculated in \cite{Herzog:2006gh}.}. 
It is of interest to see how the various 
quantities change if we consider Yang-Mills theories in other dimensions 
which are non-conformal\footnote{Non-conformal theories have also been 
considered, among other things, in \cite{Caceres:2006ta} and we thank Makoto
Natsuume for bringing this reference to our attention.}. 
So, in this paper we start from the non-extremal
D$p$-brane solution \cite{Horowitz:1991cd}, a particular decoupling limit
\cite{Itzhaki:1998dd} of which defines the
gravity dual of the $(p+1)$-dimensional SU($N$) Yang-Mills theory at 
large $N$. We then apply
the fundamental string probe approach and compute the Nambu-Goto world-sheet 
action for this background. The expectation value of the required Wilson loop
corresponds to the above minimal area whose boundary is the loop in question
\cite{Maldacena:1998im}.  
We consider both the time-like as well as light-like Wilson loops.  
We first compute the time-like Wilson loop when the velocity
of the dipole is arbitrary but less than 1. From there we obtain the 
quark-antiquark potential of a dipole moving through the $(p+1)$-dimensional 
Yang-Mills plasma by performing numerical integration. This gives us exact
quark-antiquark potential at different values of its velocity. This was known
previously for $p=3$ in \cite{Liu:2006he}, but here we obtain in addition 
the results for $p=2,4$ and 5 as well. We have also plotted both the
quark-antiquark separation and the potential for various values of $p$
at a fixed velocity to see the differences. 
Next we compute the screening length of the dipole not
only at the leading order (as obtained in \cite{Caceres:2006ta}), 
but also
at the higher order in velocity and give their analytic expressions. Higher 
order results were known only for
$p=3$ in \cite{Liu:2006he} and the leading order in other $p$'s in
\cite{Caceres:2006ta} (the leading order results of the screening lengths
for general $p$ were first obtained in this paper), but here we 
calculate the higher order corrections in
screening lengths for $p=2$ and 4. We have given the results for $p=3$ also
for comparison. Then we calculate the jet quenching parameter from 
the light-like Wilson loop, i.e. by taking the velocity going to 1 limit of
the previous calculation. 
Our calculation is a careful rederivation of the jet quenching parameter by 
the method used in \cite{Liu:2006he} for $p=3$ applied to other $p$'s. 
              
This paper is organized as follows. In section 2 we compute the
time-like Wilson loop and from there obtain the quark-antiquark potential as
well as the screening length of the dipole moving through the 
$(p+1)$-dimensional Yang-Mills plasma with an arbitrary velocity. In section
3, we give the derivation of the jet quenching parameter from the light-like
Wilson loop. Finally, we conclude in section 4.

\section{The $q$-${\bar q}$ potential and the screening
length}

Using AdS/CFT correspondence, we calculate in this section the expectation 
value of the time-like Wilson loop of the $(p+1)$-dimensional Yang-Mills theory
by calculating the Nambu-Goto action of a fundamental string in the background
of a non-extremal D$p$-brane in a particular decoupling limit. From this
we will obtain the velocity dependent quark-antiquark potential and the
screening length of the dipole.
 
The metric (given in the string frame), the dilaton and the form-field of
the non-extremal D$p$-brane solution of type II supergravity are given as
\cite{Horowitz:1991cd},
\bea\label{dpbrane}
ds^2 &=& H^{-\frac{1}{2}}\left(-f dt^2 + \sum_{i=1}^p(dx^i)^2\right)
+ H^{\frac{1}{2}}\left(\frac{dr^2}{f} + r^2 d\Omega_{8-p}^2\right)\nn
e^{2(\phi-\phi_0)} &=& H^{\frac{3-p}{2}},\qquad F_{[p+2]} = \coth\alpha\,
dH^{-1}\wedge dt \wedge dx^1\wedge \ldots \wedge dx^p
\eea   
Here the functions $H(r)$ and $f(r)$ are defined as,
\be
H(r) = 1 + \frac{r_0^{7-p} \sinh^2\alpha}{r^{7-p}}, \qquad f(r) = 1 - 
\frac{r_0^{7-p}}{r^{7-p}}
\ee
where $r_0$ and $\alpha$ are two parameters related to the mass and the charge
of the black D$p$-brane. There is an event horizon at $r=r_0$ and $e^{\phi_0}=
g_s$ is the string coupling constant. The form-field $F_{[p+2]}$ has to be
made self-dual for $p=3$. In the decoupling limit we zoom into the
region, 
\be
r_0^{7-p} < r^{7-p} \ll r_0^{7-p}\sinh^2\alpha
\ee
So, $\alpha$ is a very large angle and we can neglect 1 in $H(r)$, i.e.,
\be
H(r) \approx \frac{r_0^{7-p} \sinh^2\alpha}{r^{7-p}}
\ee     
and the metric now takes the form,
\be\label{ymmetric}
ds^2 = \frac{r^{\frac{7-p}{2}}}{r_0^{\frac{7-p}{2}}\sinh\alpha}\left(-f dt^2
+ \sum_{i=1}^p (dx^i)^2\right) +
\frac{r_0^{\frac{7-p}{2}}\sinh\alpha}{r^{\frac{7-p}{2}}} \frac{dr^2}{f} + 
\frac{r_0^{\frac{7-p}{2}}\sinh\alpha}{r^{\frac{3-p}{2}}}d\Omega_{8-p}^2
\ee
Along with the other field configurations this is the gravity dual of
$(p+1)$-dimensional finite temperature SU($N$) Yang-Mills theory 
\cite{Itzhaki:1998dd}. We use 
open string as a probe and consider its dynamics in this background. Let the
line joining the end points of the open string, i.e., the dipole lie along 
$x^1$-direction and move with an arbitrary velocity $0<v<1$ along 
$x^p$-direction. Since the dipole lies perpendicular to its direction of
propagation, so $p$ must be greater than 1. Now we can go 
to the rest frame $(t',\, x^p\,')$ of the
quark-antiquark by boosting the coordinate system as,
\bea\label{boost}
dt &=&  \cosh\eta\,dt' - \sinh\eta\,(dx^p)'\nn
dx^p &=& -\sinh\eta\,dt' + \cosh\eta\,(dx^p)'
\eea
where the boost parameter $\eta$ is related to $v$ as $\tanh\eta = v$. In this
frame the dipole is static and the quark-gluon plasma is moving
with velocity $v$ in the negative $x^p$-direction. The Wilson loop lies in the 
$t'$-$x^1\,'$ plane and we denote the lengths as ${\cal T}$ and $L$ in those
directions. We further assume ${\cal T} \gg L$ such that the string
world-sheet is time translation invariant. Using \eqref{boost} in the metric
\eqref{ymmetric} we get,
\bea\label{newframe}
ds^2 &=& -A(r) dt^2 - 2B(r)dtdx^p + C(r) (dx^p)^2 + 
\frac{r^{\frac{7-p}{2}}}{r_0^{\frac{7-p}{2}}\sinh\alpha}
\sum_{i=1}^{p-1} (dx^i)^2\nn 
& & \qquad\qquad\qquad\qquad\qquad +
\frac{r_0^{\frac{7-p}{2}}\sinh\alpha}{r^{\frac{7-p}{2}}} \frac{dr^2}{f} + 
\frac{r_0^{\frac{7-p}{2}}\sinh\alpha}{r^{\frac{3-p}{2}}}d\Omega_{8-p}^2\nn
&=& G_{\mu\nu} dx^\mu dx^\nu
\eea
where
\bea\label{metricfns}
A(r) &=& \frac{r^{\frac{7-p}{2}}}{r_0^{\frac{7-p}{2}}\sinh\alpha}\left(1 -
\frac{r_0^{7-p} \cosh^2\eta}{r^{7-p}}\right)\nn
B(r) &=& \frac{r_0^{\frac{7-p}{2}}}{r^{\frac{7-p}{2}}\sinh\alpha} 
\sinh\eta \cosh\eta\nn
C(r) &=& \frac{r^{\frac{7-p}{2}}}{r_0^{\frac{7-p}{2}}\sinh\alpha}\left(1 +
\frac{r_0^{7-p} \sinh^2\eta}{r^{7-p}}\right) 
\eea
Also note that since we will be using the primed coordinates from now on, we
have dropped the `prime' in writing \eqref{newframe} for brevity. 
We will evaluate
the world-sheet Nambu-Goto action given by,
\be\label{ngaction}
S =  \frac{1}{2\pi\alpha'}\int d\sigma d\tau \sqrt{-{\rm det} g_{\alpha\beta}}
\ee
in this background. Here $g_{\alpha\beta}$ is the induced metric on the 
world-sheet
\be
g_{\alpha\beta} = G_{\mu\nu}\frac{\partial x^\mu}{\partial \xi^{\alpha}}
\frac{\partial x^\nu}{\partial \xi^{\beta}}
\ee
with $\xi^{\alpha} = \tau,\,\sigma$ for $\alpha = 0,1$ respectively. We choose
the static gauge condition for evaluating \eqref{ngaction} as, $\tau = t$,
$\sigma = x^1$, where $-L/2 \leq x^1 \leq L/2$ and $r=r(\sigma)$, $x^2(\sigma)
= x^3(\sigma) = \cdots = x^p(\sigma) =$ constant. $r(\sigma)$ is the string
embedding we want to determine with the boundary condition, 
$r(\pm \frac{L}{2}) = r_0\Lambda$. Using these in \eqref{ngaction}, we get
\be\label{ngaction1}
S = \frac{{\cal T}}{2\pi\alpha'}\int_{-L/2}^{L/2} d\sigma \left[A(r)\left(
\frac{r^{\frac{7-p}{2}}}{r_0^{\frac{7-p}{2}}\sinh\alpha} + 
\frac{r_0^{\frac{7-p}{2}}\sinh\alpha}{r^{\frac{7-p}{2}}}
\frac{(\partial_{\sigma}r)^2}{f}\right)\right]^{\frac{1}{2}}
\ee
Now defining new dimensionless variables $y=r/r_0$, and also $\tilde \sigma =
\sigma/(r_0\sinh\alpha)$, $\ell = L/(r_0\sinh\alpha) = 
4\pi L T/(7-p)$, where
$T$ is the Hawking temperature that can be obtained from the non-extremal
D$p$-brane metric in \eqref{dpbrane} as $T=(7-p)/(4\pi r_0\sinh\alpha)$, the
action \eqref{ngaction1} reduces to,
\be\label{ngaction2}
S = \frac{{\cal T}r_0}{\pi\alpha'}\int_0^{\ell/2}d\tilde \sigma {\cal L}
=  \frac{{\cal T} d_p^{\frac{1}{5-p}}\lambda^{\frac{1}{5-p}}(4\pi
  T)^{\frac{2}{5-p}}}{\pi (7-p)^{\frac{2}{5-p}}}\int_0^{\ell/2}d\sigma {\cal L}
\ee 
where 
\be\label{lagrangian}
{\cal L} = \sqrt{\left(y^{7-p} -
    \cosh^2\eta\right)\left(1+\frac{y'^2}{y^{7-p}-1}\right)}
\ee
with $y'=\partial y/\partial \sigma$. Here we have used the fact that $y$ is 
an even fuction of $\sigma$ by symmetry. Note that in writing the second
expression in \eqref{ngaction2}, we have used the standard formulae
\cite{Itzhaki:1998dd},
\bea\label{parameters}
r_0^{7-p} \sinh^2\alpha &=& d_p g_{\rm YM}^2 N \alpha'^{5-p} = d_p \lambda
\alpha'^{5-p} \nn
r_0 \sinh\alpha &=& \frac{7-p}{4\pi T}
\eea
where $d_p = 2^{7-2p}\pi^{(9-3p)/2}\Gamma((7-p)/2)$ and $\lambda = g_{\rm
  YM}^2 N$, the 't Hooft coupling, $N$ being the number of D$p$-branes which 
in gauge theory is the rank of the gauge group. In the above $p$ has been
assumed to be less than 5. We will mention about $p=$ 5 and 6 later. Also 
note that in the second
expression of \eqref{ngaction2}, we have omitted the `tilde' in $\sigma$ for
brevity. $y(\sigma)$ is determined by extremizing \eqref{ngaction2}. Now since
the Lagrangian density given in \eqref{lagrangian} does not depend explicitly
on $\sigma$, we have
\be\label{com}
{\cal H} = {\cal L} - y'\frac{\partial {\cal L}}{\partial y'} = \frac{y^{7-p}
  - \cosh^2\eta}{\sqrt{\left(y^{7-p}-\cosh^2\eta\right)
\left(1+\frac{y'^2}{y^{7-p}-1}\right)}} = {\rm const.}
\ee         
As explained in \cite{Liu:2006he} for D3-brane, we will consider two 
cases: (i) In this case,
$\cosh^{\frac{2}{7-p}}\eta < \Lambda$ and then take $\Lambda \to \infty$. So,
the rapidity $\eta$ remains finite. The Wilson loop in this case is time-like
and the action is real. We will compute the quark-antiquark potential and the
screening length for this case in this section. (ii) In this case,  
initially we take
$\cosh^{\frac{2}{7-p}}\eta > \Lambda$ and then take $\eta \to \infty$ keeping
$\Lambda$ finite. The Wilson loop in this case would be light-like and the
action is imaginary. We will take $\Lambda \to \infty$ at the end and obtain
the expression for the jet quenching parameter. This will be considered in
the next section. 

For case (i) when $\cosh^{\frac{2}{7-p}}\eta < \Lambda$ and the action is
real, let us denote the constant of motion \eqref{com} as $q$. Then
$y'$ can be solved and we get from \eqref{com},
\be\label{yprime}
y' = \frac{1}{q}\sqrt{\left(y^{7-p} - 1\right)\left(y^{7-p} -
    y_c^{7-p}\right)}
\ee
where $y_c^{7-p} = \cosh^2\eta + q^2 > 1$, denotes the largest turning point
where $y'$ vanishes. Integrating 
this equation we obtain,
\be\label{lq}
2\int_0^{\ell/2} d\sigma = \ell(q) = 2q\int_{y_c}^\infty \frac{dy}{
\sqrt{\left(y^{7-p}-1\right)\left(y^{7-p} - y_c^{7-p}\right)}}
\ee
Note here that we have taken the boundary $\Lambda \to \infty$.
Eq.\eqref{lq} therefore gives us the separation between the quark and the 
antiquark in the dipole as a function of the integration constant $q$.
The integral expression for the quark-antiquark separation for the general
metric including D$p$-branes has been given in \cite{Caceres:2006ta}.  
It is difficult to integrate the expression on the rhs of \eqref{lq} and
write an analytic expression for $\ell(q)$ in general. However, we can
give analytic expression for large rapidity $\eta$ or large $y_c$ and from
there we can obtain the form of screening length which will be discussed 
later.

Now substituting the form of $y'$ from \eqref{yprime} into the action
\eqref{ngaction2} along with \eqref{lagrangian} and changing the variable
from $\sigma$ to $y$, we get,
\be\label{ngaction3}
S(\ell) =  \frac{{\cal T} d_p^{\frac{1}{5-p}}\lambda^{\frac{1}{5-p}}(4\pi
  T)^{\frac{2}{5-p}}}{\pi (7-p)^{\frac{2}{5-p}}}\int_{y_c}^{\infty}
dy \frac{y^{7-p} - \cosh^2\eta}{\sqrt{\left(y^{7-p} -1\right)
\left(y^{7-p} - y_c^{7-p}\right)}}
\ee
Note that here we have expressed $S$ completely in terms of the parameters
of the gauge theory. In order to calculate the quark-antiquark potential
we must subtract from it the quark and antiquark self-energy $S_0$. 
If $E(L)$ is the potential then,
\be\label{potential}
E(L) = \frac{S(\ell) - S_0}{\cal T}
\ee

Now to compute $S_0$, we consider an open string along radial direction, i.e. 
a single quark in the same background \eqref{newframe} as before and use the
static gauge condition $\tau = t$, $\sigma = r$, $x^p = x^p(\sigma)$ and
$x^1(\sigma) = x^2(\sigma) = \cdots = x^{p-1}(\sigma) =$ constant. With these
we evaluate the Nambu-Goto world-sheet action and then multiply by 2 to get the
contribution for two strings. From \eqref{ngaction} we get in this case,
\be\label{ngaction0}
S_0 = \frac{2{\cal T}}{2\pi\alpha'}\int_{r_0}^{\infty} dr
\sqrt{\frac{r_0^{\frac{7-p}{2}}\sinh\alpha}{r^{\frac{7-p}{2}}}
\frac{A(r)}{f}+\left(A(r)C(r)+B(r)^2\right) (x^p\,')^2}
\ee
where $A(r)$, $B(r)$ and $C(r)$ are as given before in \eqref{metricfns}. 
Note here that the string stretches all the way upto the horizon $r_0$. Now
introducing new dimensionless variables as before $y = r/r_0$ 
and $z=x^p/(r_0\sinh\alpha)$
and substituting $r_0/\alpha'$ in terms of the parameters of the gauge theory
we get from \eqref{ngaction0},
\be\label{ngaction01} 
S_0 =   \frac{{\cal T} d_p^{\frac{1}{5-p}}\lambda^{\frac{1}{5-p}}(4\pi
  T)^{\frac{2}{5-p}}}{\pi (7-p)^{\frac{2}{5-p}}}\int_1^{\infty} dy
\sqrt{\frac{y^{7-p}-\cosh^2\eta}{y^{7-p} - 1} + \left(y^{7-p} - 1\right)
\left(\frac{\partial z}{\partial y}\right)^2}
\ee
Since the Lagrangian density in \eqref{ngaction01} is independent of $z$, the
Euler-Lagrange equation of motion gives a conservation relation $(\partial{\cal
  L}/\partial(\partial_y z)) =$ const. independent of $y$. Denoting the
constant by $\tilde q$, we get from this,
\be\label{zprime}
\left(\frac{\partial z}{\partial y}\right)^2 = \tilde q^2\frac{y^{7-p} -
  \cosh^2\eta }{\left(y^{7-p} - 1\right)^2\left(y^{7-p} - \tilde q^2
    -1\right)}
\ee   
Since $y$ varies from 1 to $\infty$, the right hand side can become negative
and unphysical for arbitrary values of $\eta$ and $\tilde q$. So, in order to
get physical solution we must choose the constant $\tilde q =
\sinh\eta$. Therefore, we get
\be\label{zprime1}
\frac{\partial z}{\partial y} = \frac{\sinh\eta}{\left(y^{7-p} - 1\right)}
\quad \Rightarrow \quad z(y) = {\rm const.} - y \sinh\eta\,\,
_2F_1\left(1,\frac{1}{7-p}, \frac{8-p}{7-p};y^{7-p}\right)
\ee
where $_2F_1$ is the hypergeometric function. Now substituting $\partial
z/\partial y$ into \eqref{ngaction01} we get
\be
S_0 =   \frac{{\cal T} d_p^{\frac{1}{5-p}}\lambda^{\frac{1}{5-p}}(4\pi
  T)^{\frac{2}{5-p}}}{\pi (7-p)^{\frac{2}{5-p}}}\int_1^{\infty} dy
\ee
So, the quark-antiquark potential \eqref{potential} has the form,
\be\label{potential1}
E(\ell)= \frac{d_p^{\frac{1}{5-p}}\lambda^{\frac{1}{5-p}}(4\pi
  T)^{\frac{2}{5-p}}}{\pi (7-p)^{\frac{2}{5-p}}}\left[\int_{y_c}^{\infty} 
dy\left(\frac{y^{7-p} - \cosh^2\eta}
{\sqrt{(y^{7-p} - 1)(y^{7-p} - y_c^{7-p})}} - 1\right) - (y_c -1)\right]
\ee

\begin{figure} [ht]
\begin{center}
\subfigure[]{
\includegraphics[scale=0.3, angle=-90]{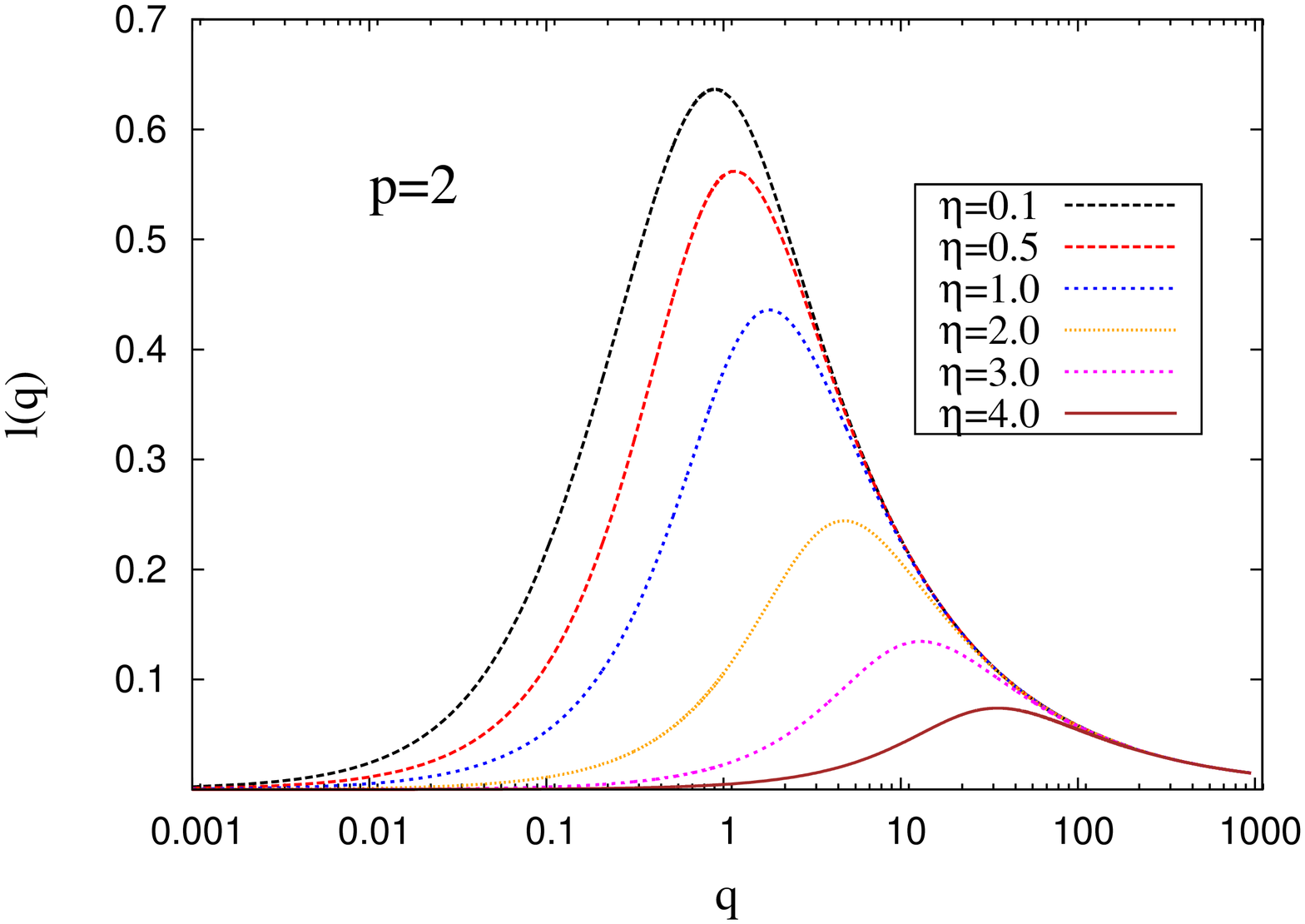}}
\subfigure[]{
\includegraphics[scale=0.3, angle=-90]{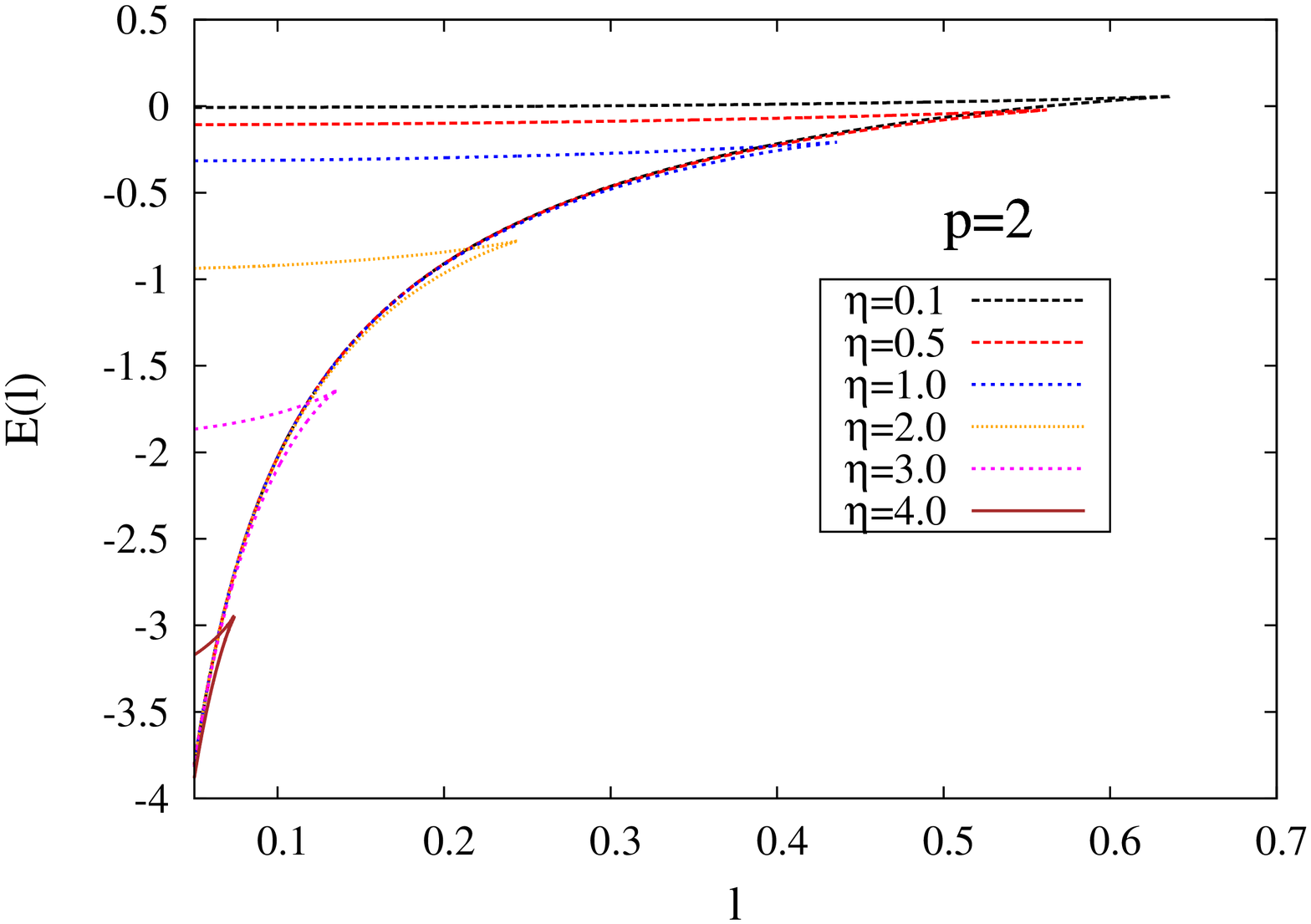}}
\caption{\label{fig:qm/complefunctions} \small{(a) shows the plot of 
quark-antiquark
separation $\ell$ as a function of integration const. $q$ for $p=2$ at
different rapidities $\eta$ of the dipole. (b) shows the plot
of properly normalized quark-antiquark potential as a function of $\ell$ for
$p=2$ at the same set of rapidities.}}
\end{center}
\end{figure}
It is in general not possible to perform the integration on the rhs of
\eqref{potential1} and obtain an analytic expression for quark-antiquark
potential $E(\ell)$. So, as in \cite{Liu:2006he,Liu:2006nn}, we will first 
plot $\ell(q)$ vs 
$q$ for certain particular values of $\eta$ from the
integral equation \eqref{lq} and obtain $q$ as a function of $\ell$ and then
using these $q$ in the integral equation \eqref{potential1} we plot $E(\ell)$
vs $\ell$ for those values of $\eta$. In \cite{Liu:2006he,Liu:2006nn}, these 
plots were given for
$p=3$, we here give the plots for $p=2,\,4$ and 5 in Figures 1, 2 and 3
respectively. Also for comparison with different $p$'s (including $p=3$) we 
give the plot of both $\ell(q)$ vs $q$ and $E(\ell)$ vs $\ell$ in Figure 4
at $\eta=1$. 

We have mentioned before that we are mainly considering the cases
with $p < 5$. This is because the constant (expressed in terms of the
parameters of the gauge theory by \eqref{parameters}) in front of the second 
expression in \eqref{ngaction2} is ill defined for $p=5$. But no such problem
arises if we keep the parameters $r_0$ and $\alpha'$ as in the first 
expression in \eqref{ngaction2} of the gravity theory. In fact we see
from \eqref{parameters} that for $p=5$ we can not express $r_0/\alpha'$ in
terms of the parameters of the gauge theory. This may be an indication that in
this case the complete decoupling does not occur. However, we can still plot
$\ell(q)$ vs $q$ and $E(\ell)$ vs $\ell$, as we do for $p=5$, keeping the
constant in terms of the parameters of the gravity side. Even for $p=6$ case,
it is known that the decoupling does not occur and so, we do not plot the
functions.
\begin{figure} [ht]
\begin{center}
\subfigure[]{
\includegraphics[scale=0.3, angle=-90]{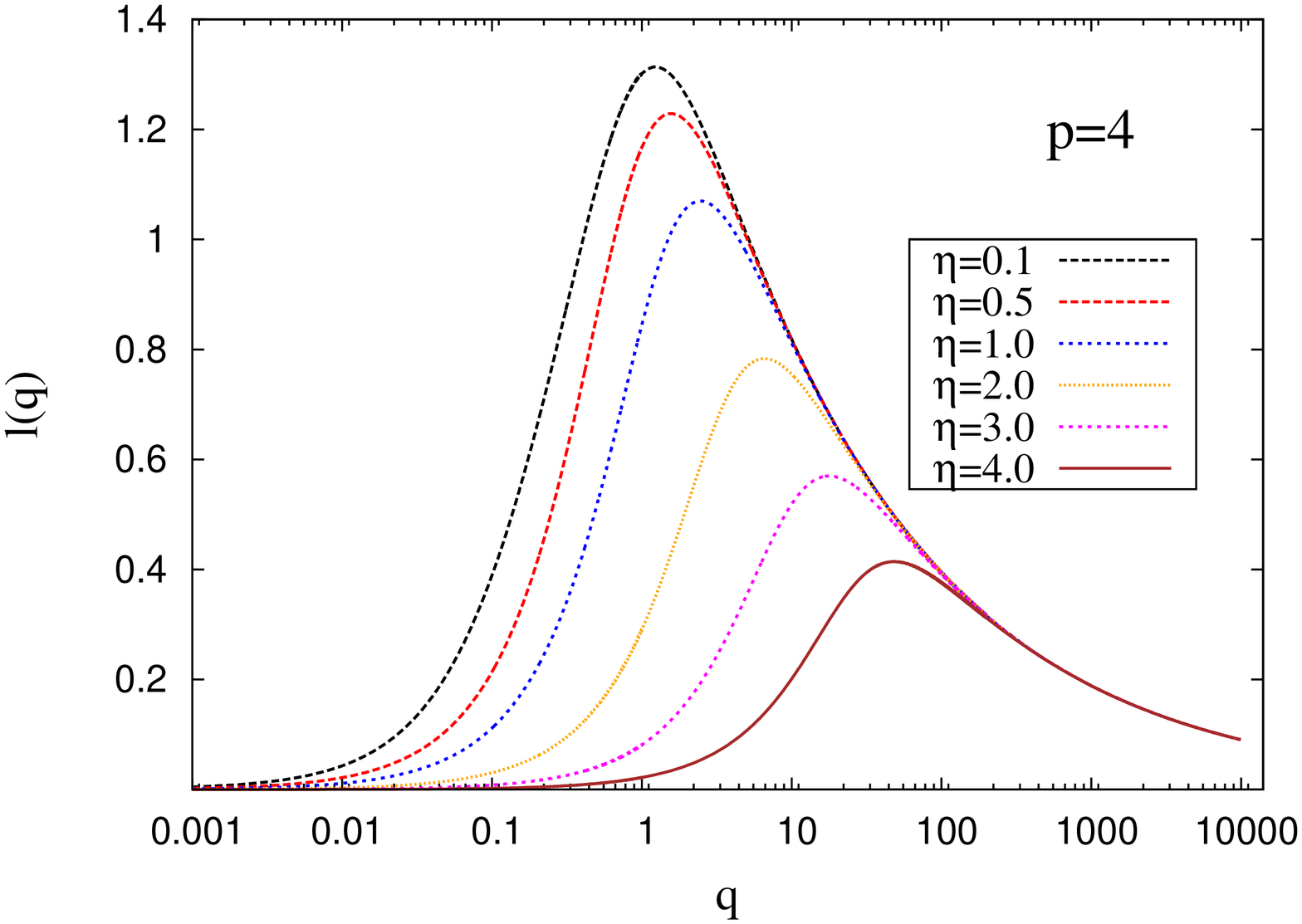}}
\subfigure[]{
\includegraphics[scale=0.3, angle=-90]{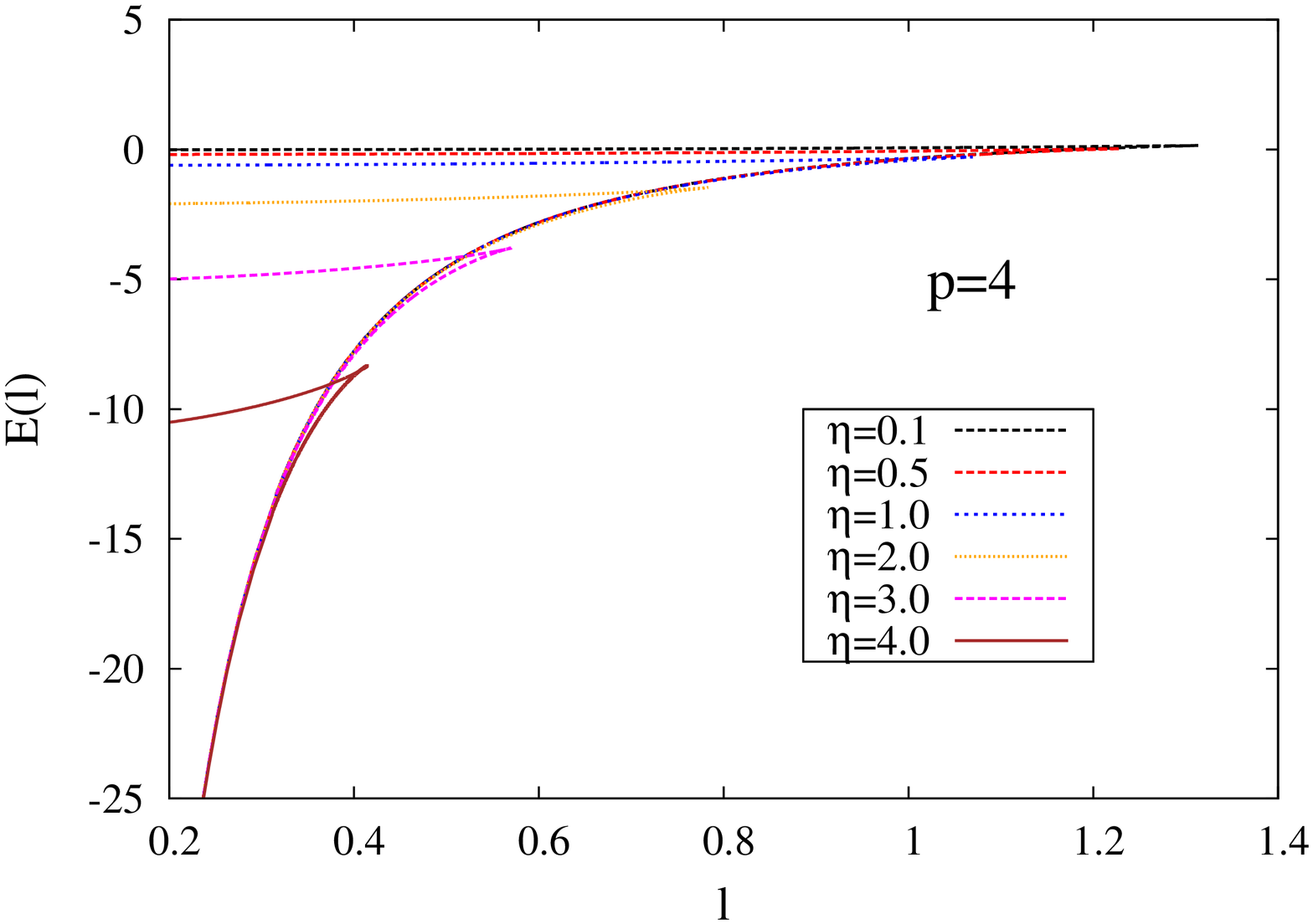}}
\caption{\label{fig:qm/complexfuncts} \small{(a) shows the plot of 
quark-antiquark
separation $\ell$ as a function of integration const. $q$ for $p=4$ at
different rapidities $\eta$ of the dipole. (b) shows the plot
of properly normalized quark-antiquark potential as a function of $\ell$ for
$p=4$ at the same set of rapidities.}}
\end{center}
\end{figure}
\begin{figure} [ht]
\begin{center}
\subfigure[]{
\includegraphics[scale=0.3, angle=-90]{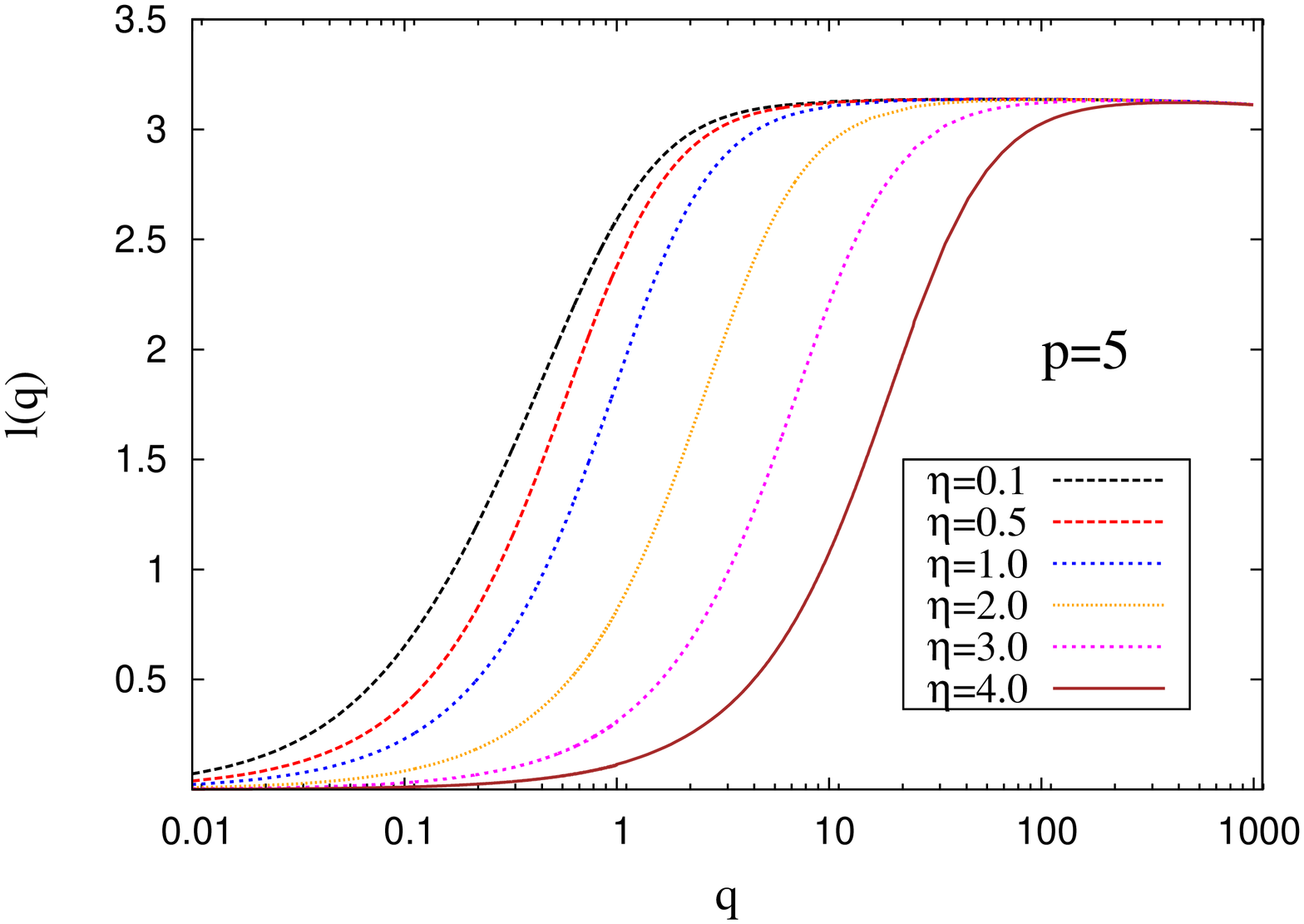}}
\subfigure[]{
\includegraphics[scale=0.3, angle=-90]{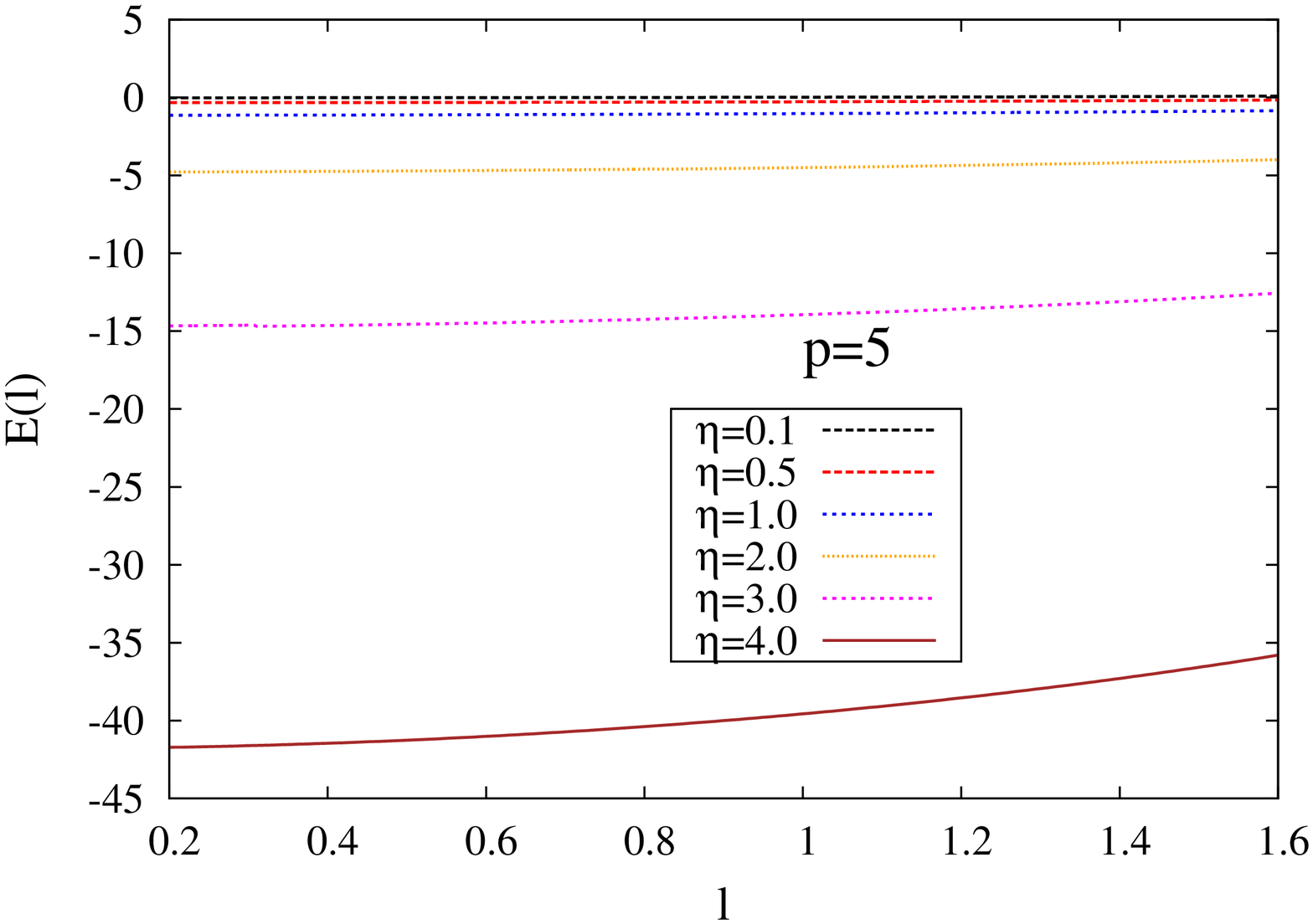}}
\caption{\label{fig:qm/complexfunctis}\small{(a) shows the plot of 
quark-antiquark
separation $\ell$ as a function of integration const. $q$ for $p=5$ at
different rapidities $\eta$ of the dipole. Here $\ell$ saturates 
unlike in Figures 1 and 2. (b) shows the plot
of properly normalized quark-antiquark potential as a function of $\ell$ for
$p=5$ at the same set of rapidities. There is no lower branch unlike in Figures
1 and 2.}} 
\end{center}
\end{figure}
\begin{figure} [ht]
\begin{center}
\subfigure[]{
\includegraphics[scale=0.3, angle=-90]{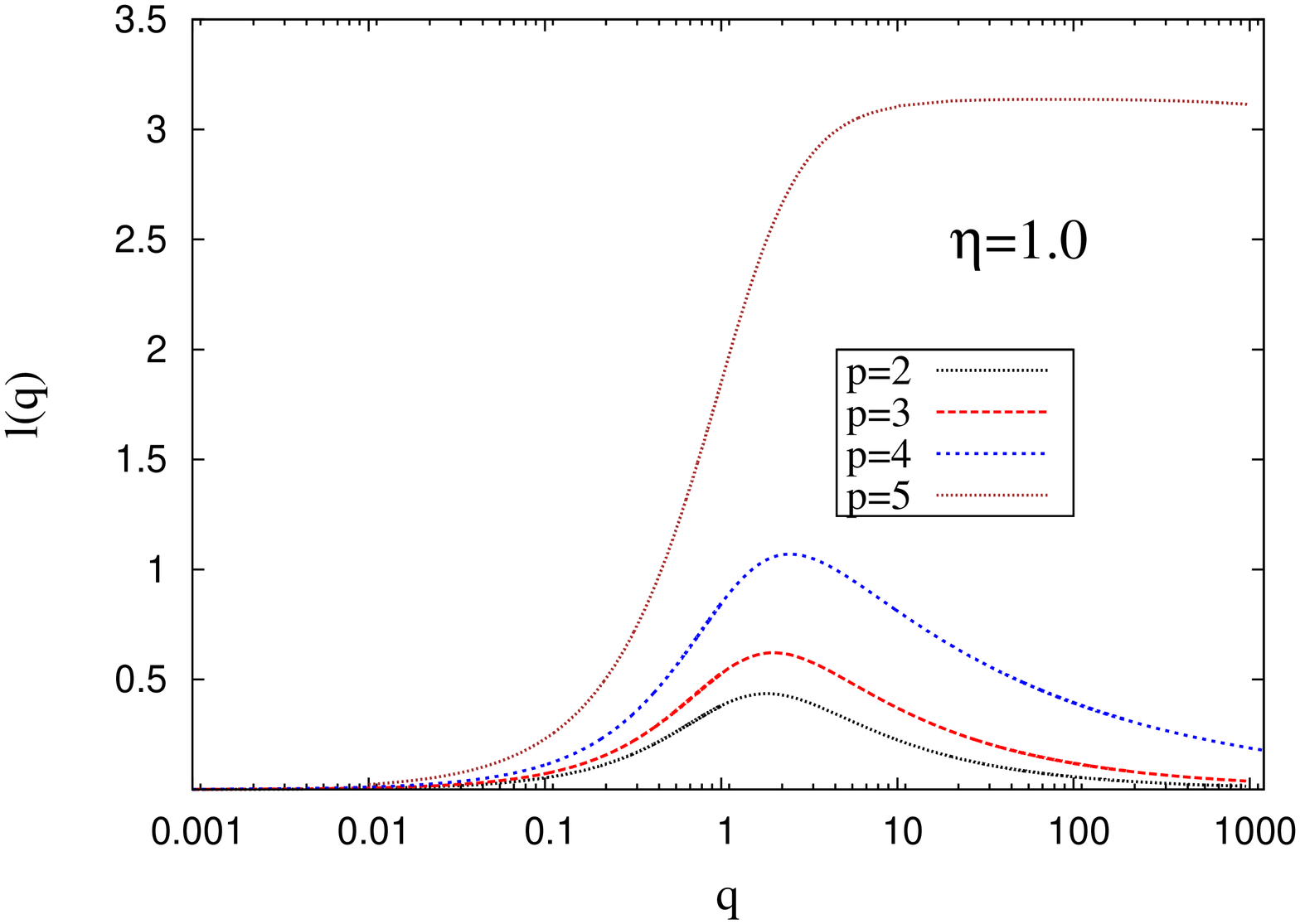}}
\subfigure[]{
\includegraphics[scale=0.3, angle=-90]{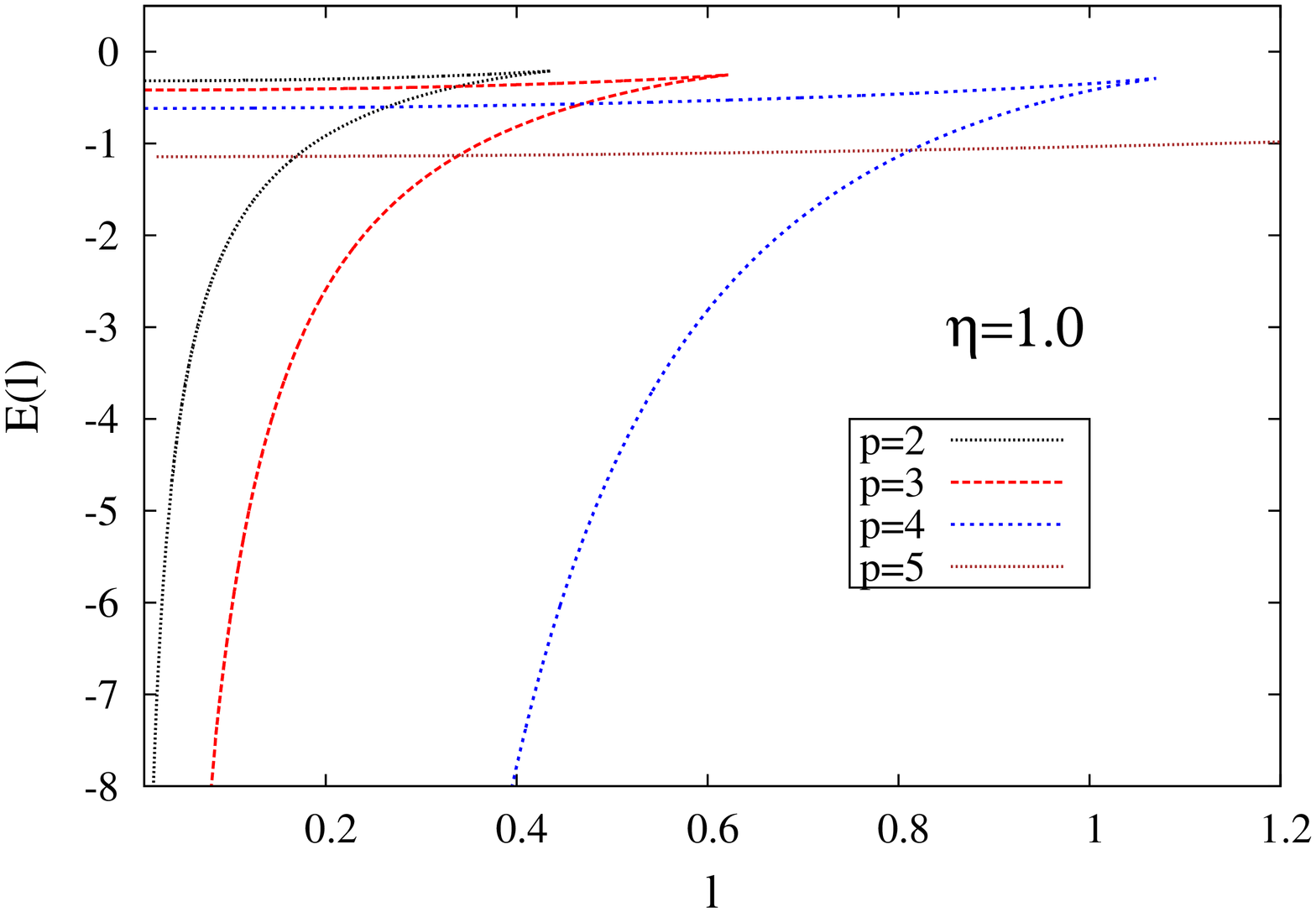}}
\caption{\label{fig:qm/complexfunction} \small{(a) shows the plot of 
quark-antiquark
separation $\ell$ as a function of integration const. $q$ for different values
of $p$ for comparison at $\eta = 1.0$. (b) shows the plot of quark-antiquark
potential as a function of $\ell$ for different values of $p$ for comparison
at the same $\eta=1.0$.}}
\end{center}
\end{figure}

The general features of the plot for $p=2,\,4$ remain very similar (although
the details, as shown in Figure 4 below, are quite different) to $p=3$
case discussed in \cite{Liu:2006he,Liu:2006nn}. It is clear from 
\eqref{lq} that $\ell(q)$ goes to
zero as $q$ for small $q$ (for all $p$) and as $q^{-(5-p)/(7-p)}$ for 
large $q$ (for $p<5$). However, for $p=5$, it goes to a constant for large 
$q$. These can be seen in Figures 1, 2, 3. Also, for $p<5$,
the plots show that it has a maximum $\ell_{\rm max}$ in between. Beyond this
there is no solution of \eqref{lq}. From Figures 1, 2 we see that the 
peak of the $\ell(q)$ 
curve reduces and shifts towards right, i.e, towards a larger value of $q$ 
as we 
increase $\eta$ or the rapidity. From Figure 4(a), we see that at a fixed 
value of $\eta$, the peak reduces as we increase $p$ and shifts towards left
i.e., towards a lower value of $q$. As $\ell(q)$ decreases from $\ell_{\rm
  max}$, there are two dipoles at a fixed $\ell$ for two different values of 
$q$.
The quark-antiquark potential in general decreases with increasing values
of $\eta$ at each $p$ and has two branches corresponding to the two values of
$q$. The smaller value of $q$ corresponds to the upper branch and has higher
energy, whereas the larger value of $q$ corresponds to the lower branch and has
lower energy. So, the dipole with lower $q$ will be metastable and will go to 
the state with higher $q$ as it is energetically more favorable.

Also, there exists a critical $\eta_{c}$ above which the whole upper branch
of the $E(\ell)$ curve is negative. But for $\eta < \eta_{c}$ the $E(\ell)$
curve crosses zero at $\ell=\ell_{c}$, continues to rise till 
$\ell=\ell_{\rm max}$
and turns back crossing zero again at $\ell=\ell'_{c}> \ell_{c}$. Below 
$\ell_{c}$, the 
upper branch is metastable. A dipole on the upper branch on slight perturbation
will come down to the lower branch.  At $\ell=\ell_c$,
the dipole in the upper branch and the two isolated string configurations
(or dissociated quark and antiquark) have
the same energy. So, both the states can coexist. However with slight
disturbance it will settle down to the dipole in the lower branch. 
In the regime $\ell_{c}<\ell<\ell'_{c}$ 
the upper branch has positive energy while the lower one has negative energy. 
So a dipole sitting 
on the upper branch, when perturbed, may either come down and settle 
in the lower branch
or it may dissociate into a free quark and a free antiquark. At 
$\ell=\ell'_c$, the dipole in the
lower branch and the two isolated string states (or dissociated quark and
antiquark) can coexist and both are stable configurations.
 In the domain $\ell'_{c}<\ell< \ell_{\rm max}$
both the branches have positive energy and so a dipole sitting on either 
of them will dissociate when slightly disturbed. Beyond $\ell_{\rm max}$ 
no dipole will be formed at all.

Some of these features were mentioned in \cite{Liu:2006he,Liu:2006nn} 
for $p=3$, but it continues to hold for $p=2,\,4$ cases as well.
For $p=5$, since there is no maximum for $\ell(q)$ plot, there is no lower 
branch in the $E(\ell)$ vs $\ell$ plot. The plot of quark-antiquark potential 
$E(\ell)$ for different values of $p$ are given in Figure 4(b) for comparison.

We mentioned before that $\ell(q)$ in \eqref{lq} can not be integrated in
general. However, for large $\eta$ or large $y_c$, we can expand $\ell(q)$ 
and then integrate to write a series expansion in powers of $1/y_c$ as,
\bea\label{lq1}
\ell(q) &=&
2q\int_{y_c}^{\infty}\frac{dy}{y^{\frac{7-p}{2}}
(y^{7-p}-y_c^{7-p})^\frac{1}{2}} + q \int_{y_c}^{\infty}\frac{dy}
{y^{\frac{3(7-p)}{2}}(y^{7-p}-y_c^{7-p})^\frac{1}{2}}\nn 
& & \qquad\qquad + \frac{3q}{4}
\int_{y_c}^{\infty}\frac{dy}{y^{\frac{5(7-p)}{2}} 
(y^{7-p}-y_c^{7-p})^\frac{1}{2}} + \cdots
\eea        
and on integration this yields for $p=2,\,3$ and 4,
\bea
\ell(q)^{p=2} &=& \frac{2q\sqrt{\pi}}{5y_c^4}\left[\frac{\Gamma(\frac{4}{5})}
{\Gamma(\frac{13}{10})} + \frac{\Gamma(\frac{9}{5})}{10\Gamma(\frac{23}{10})}
\frac{1}{y_c^5} +
\frac{3\Gamma(\frac{14}{5})}{8\Gamma(\frac{33}{10})}\frac{1}{y_c^{10}} +
\cdots\right]\\
\ell(q)^{p=3} &=& \frac{2q\sqrt{\pi}}{y_c^3}\left[\frac{\Gamma(\frac{3}{4})}
{\Gamma(\frac{1}{4})} + \frac{\Gamma(\frac{7}{4})}{8\Gamma(\frac{9}{4})}
\frac{1}{y_c^4} +
\frac{3\Gamma(\frac{11}{4})}{32\Gamma(\frac{13}{4})}\frac{1}{y_c^{8}} +
\cdots\right]\\
\ell(q)^{p=4} &=& \frac{4q\sqrt{\pi}}{y_c^2}\left[\frac{\Gamma(\frac{2}{3})}
{\Gamma(\frac{1}{6})} + \frac{\Gamma(\frac{5}{3})}{12\Gamma(\frac{13}{6})}
\frac{1}{y_c^3} +
\frac{\Gamma(\frac{8}{3})}{16\Gamma(\frac{19}{6})}\frac{1}{y_c^{6}} +
\cdots\right]
\eea
By truncating the series upto the second term we can calculate $\ell_{\rm
  max}$ for the above three cases as,
\bea\label{p2}
\ell_{\rm max}^{p=2} &=& \frac{2\, \cdot
  3^{3/10}\sqrt{\pi}\Gamma(\frac{4}{5})}{8^{4/5}\sqrt{5}
\Gamma(\frac{13}{10})}\left[\frac{1}{\cosh^{\frac{3}{5}}\eta} +
  \frac{3}{130} \frac{1}{\cosh^{\frac{13}{5}}\eta} + \cdots\right]\nn
&=& 0.54176\left[\frac{1}{\cosh^{\frac{3}{5}}\eta} +
  \frac{3}{130} \frac{1}{\cosh^{\frac{13}{5}}\eta} + \cdots\right]\\
\label{p3}
\ell_{\rm max}^{p=3} &=& \frac{2 \sqrt{2\pi}
  \Gamma(\frac{3}{4})}{3^{3/4}
\Gamma(\frac{1}{4})}\left[\frac{1}{\cosh^{\frac{1}{2}}\eta} +
  \frac{1}{10} \frac{1}{\cosh^{\frac{5}{2}}\eta} + \cdots\right]\nn
&=& 0.74333\left[\frac{1}{\cosh^{\frac{1}{2}}\eta} +
  \frac{1}{10} \frac{1}{\cosh^{\frac{5}{2}}\eta} + \cdots\right]
\eea
\bea
\label{p4} \ell_{\rm max}^{p=4} &=& \frac{4^{1/3}
 \sqrt{3\pi}\Gamma(\frac{2}{3})}{
\Gamma(\frac{1}{6})}\left[\frac{1}{\cosh^{\frac{1}{3}}\eta} +
  \frac{1}{14} \frac{1}{\cosh^{\frac{7}{3}}\eta} + \cdots\right]\nn
&=& 1.18553\left[\frac{1}{\cosh^{\frac{1}{3}}\eta} +
  \frac{1}{14} \frac{1}{\cosh^{\frac{7}{3}}\eta} + \cdots\right]
\eea
The quantity $L_{\rm max} = (7-p)\ell_{\rm max}/(4\pi T)$ can be thought of as
the screening length of the dipole in the medium since this is the maximum
value of $L$ beyond which we have two dissociated quark and antiquark or two
disjoint world-sheet corresponding to $E(L) = 0$. It has been pointed out in
\cite{Liu:2006nn,Liu:2006he} for $p=3$ that if we set $\eta = 0$ in the 
above result \eqref{p3} 
which was derived for large $\eta$ is not too far off from the actual result
at $\eta=0$ and so the screening length decreases with increasing velocity
according to the scaling $L^{p=3}_{\rm max}(v) \simeq 
L^{p=3}_{\rm max}(0)/\cosh^{1/2}\eta = L^{p=3}_{\rm max}(0)/\sqrt{\gamma}$,
where $\gamma = 1/\sqrt{1-v^2}$. By looking at the similarity of the behavior
of $\ell(q)$ and $E(\ell)$ for $p=2,\,4$, with $p=3$, we may conclude that
similar behavior will also hold true for $p=2,\,4$ cases as well. Then the 
velocity dependence of the screening lengths in these two cases is of the
form, 
\bea\label{screenlength}
L^{p=2}_{\rm max}(v) &\simeq& \frac{L^{p=2}_{\rm
    max}(0)}{\cosh^{\frac{3}{5}}\eta} = \frac{L^{p=2}_{\rm
    max}(0)}{\gamma^{\frac{3}{5}}}\\
L^{p=4}_{\rm max}(v) &\simeq& \frac{L^{p=4}_{\rm
    max}(0)}{\cosh^{\frac{1}{3}}\eta} = \frac{L^{p=4}_{\rm
    max}(0)}{\gamma^{\frac{1}{3}}}    
\eea
A general expression for the leading order contribution of the screening
lengths for general $p$ has been given in \cite{Caceres:2006ta}.  
This concludes our discussion on time-like Wilson loop when
$\cosh^{\frac{2}{7-p}}\eta < \Lambda$ and $\eta$ remains finite while $\Lambda
\to \infty$.  

\section{The jet quenching parameter}

So far in our discussion we assumed that the rapidity $\eta$ is finite and 
$\cosh^{\frac{2}{7-p}}\eta < \Lambda$. So,
the velocity of the string is in the range $0<v<1$ and the Wilson loop is
time-like. Now we will consider case (ii), i.e., $\cosh^{\frac{2}{7-p}}\eta >
\Lambda$. In order to extract the jet quenching parameter we 
take $\eta \to
\infty$ or $v \to 1$, so that the Wilson loop is light-like and then
take $\Lambda \to \infty$. 
(We will be brief here since the jet quenching parameter for 
$(p+1)$-dimensional
Yang-Mills theory has already been given in \cite{Liu:2006he,Liu:2006ug}. But 
here we obtain it
by taking $v \to 1$ limit of the time-like Wilson loop at  $0<v<1$ as was
done there for $p=3$ case.)  
Note from \eqref{ngaction2} that since
now $\cosh^{\frac{2}{7-p}}\eta > \Lambda$, the action is imaginary and we
write the second expression in \eqref{ngaction2} as,
\be\label{ngactionjet}
S =  i\frac{{\cal T} d_p^{\frac{1}{5-p}}\lambda^{\frac{1}{5-p}}(4\pi
  T)^{\frac{2}{5-p}}}{\pi (7-p)^{\frac{2}{5-p}}}\int_0^{\ell/2}d\sigma {\cal L}
\ee 
where 
\be\label{lagrangianjet}
{\cal L} = \sqrt{\left(\cosh^2\eta - y^{7-p}\right)
\left(1+\frac{y'^2}{y^{7-p}-1}\right)}
\ee     
As before since the Lagrangian density \eqref{lagrangianjet} does not
explicitly depend on $\sigma$, the corresponding Hamiltonian is conserved. So,
we have,
\be\label{conserved}
{\cal H} = {\cal L} - y'\frac{\partial{\cal L}}{\partial y'} = {\rm const.}
\quad \Rightarrow \quad \frac{\cosh^2\eta - y^{7-p}}{\sqrt{(\cosh^2\eta -
    y^{7-p}) \left(1+ \frac{y'^2}{y^{7-p}-1}\right)}} = q_0
\ee
where we have denoted the constant as $q_0$. The equation \eqref{conserved} 
can be solved for $y'$ as,
\be\label{yprimejet}
y' = \frac{1}{q_0} \sqrt{(y^{7-p} -1)(y_m^{7-p} - y^{7-p})}
\ee
where $y_m^{7-p} = \cosh^2\eta - q_0^2$. On integration eq.\eqref{yprimejet} 
gives us,
\be\label{lqjet}
\ell = 2q_0\int_1^{\Lambda} \frac{dy}{\sqrt{(y_m^{7-p} - y^{7-p})(y^{7-p}-1)}}
\ee
Substituting the value of $y'$ from \eqref{yprimejet} into the action 
\eqref{ngactionjet} we get,
\be\label{ngactionjet1}
S(\ell) = i\frac{{\cal T} d_p^{\frac{1}{5-p}}\lambda^{\frac{1}{5-p}}(4\pi
  T)^{\frac{2}{5-p}}}{\pi (7-p)^{\frac{2}{5-p}}}\int_1^{\Lambda} dy
\frac{\cosh^2\eta - y^{7-p}}{\sqrt{(y_m^{7-p} - y^{7-p})(y^{7-p}-1)}}
\ee
So far we have used only $\cosh^{\frac{2}{7-p}}\eta > \Lambda$. Now for large
$\eta$, $\ell$ in \eqref{lqjet} can be expanded as follows,
\be\label{lqjet1}
\ell = \frac{2q_0}{\cosh\eta}\int_1^\Lambda \frac{dy}{\sqrt{y^{7-p}-1}} +
O(\frac{q_0^3}{\cosh^3\eta}, \frac{\Lambda^{7-p}}{\cosh^3\eta})
\ee
Next, as we take $\eta \to \infty$ the second term in \eqref{lqjet1} drops out 
and then taking $\Lambda \to \infty$ we get,
\be\label{lqjet2}
\ell = \frac{2q_0}{\cosh\eta} a_p, \qquad {\rm with,} \quad a_p = \frac{2}{5-p}
\sqrt{\pi}
\frac{\Gamma\left(1+\frac{5-p}{2(7-p)}\right)}{\Gamma\left(\frac{6-p}{7-p}
\right)}
\ee
Further, since $L$ is much smaller than the other length dimensions of the 
problem
$\ell = (4\pi LT)/(7-p)\ll 1$ and therefore $q_0 = (\ell \cosh\eta)/(2a_p)\ll
1$. In this limit, $S(\ell)$ in \eqref{ngactionjet1} can be expanded as,
\be\label{sexpansion}
S(\ell) = S^{(0)} + q_0^2 S^{(1)} + O(q_0^4)
\ee     
where
\bea\label{relation}
S^{(0)} &=&  i\frac{{\cal T} d_p^{\frac{1}{5-p}}\lambda^{\frac{1}{5-p}}(4\pi
  T)^{\frac{2}{5-p}}}{\pi (7-p)^{\frac{2}{5-p}}}\int_1^{\Lambda} dy
\frac{\cosh^2\eta - y^{7-p}}{\sqrt{y^{7-p}-1}}\\\label{relationa}
q_0^2 S^{(1)} &=&  i\frac{{\cal T} d_p^{\frac{1}{5-p}}\lambda^{\frac{1}{5-p}}
(4\pi T)^{\frac{2}{5-p}}}{\pi (7-p)^{\frac{2}{5-p}}} q_0^2 
\int_1^{\Lambda} \frac{dy}
{\sqrt{(y^{7-p}-1)(\cosh^2\eta - y^{7-p})}}\nn
&\simeq& i\frac{({\cal T} \cosh\eta) d_p^{\frac{1}{5-p}}\lambda^{\frac{1}{5-p}}
L^2}{8\pi a_p}\left(\frac{4\pi T}{7-p}\right)^{\frac{2(6-p)}{5-p}}
\eea
Here we have used the relations $q_0 = (\ell\cosh\eta)/(2a_p)$ and $\ell =
(4\pi LT)/(7-p)$. From physical expectation it has been argued in 
\cite{Liu:2006he} that
as $\ell$ or $q_0$  goes to zero, $S^{(0)}$ is the self-energy of the two
dissociated quark and antiquark or area of the two disjoint world-sheet.  
${\cal T} \cosh\eta$ in \eqref{relationa} can be identified as
$L^-/\sqrt{2}$, where $L^-$ is the length of the Wilson loop in the 
light-like direction. Also we use the relation 
\be\label{relation1}
\langle W({\cal C})\rangle = e^{2i\left(S({\cal C}) - S_0\right)}
\approx e^{-\frac{1}{4\sqrt{2}}\hat q L^- L^2}
\ee
where the factor 2 in the exponent in the second expression is due to the fact
that we are dealing with adjoint Wilson loop. The third expression is valid
for $L \ll 1$ and also $\hat q$ is the jet quenching parameter. Thus from
\eqref{relation1} and using \eqref{relationa} we extract the value of the jet
quenching parameter as,
\be\label{jetq}
\hat q = -i\frac{8 \sqrt{2}\left(S(\ell)-S^{(0)}\right)}{L^- L^2} =
 \frac{d_p^{\frac{1}{5-p}}\lambda^{\frac{1}{5-p}}}{\pi a_p}
\left(\frac{4\pi T}{7-p}\right)^{\frac{2(6-p)}{5-p}}
\ee 
Substituting the explicit value of $a_p$ and $d_p$ given earlier it takes the
form,
\be\label{jetq1}
\hat q = \frac{4T^2\left[2^{7-2p}\pi^{\frac{9-3p}{2}}\Gamma\left(\frac{7-p}{2}
\right)\right]^{\frac{1}{5-p}} (4\pi)^{\frac{7-p}{5-p}}
\Gamma\left(\frac{6-p}{7-p}\right)}{\sqrt{\pi}\Gamma\left(\frac{5-p}{14-2p}
\right)(7-p)^{\frac{7-p}{5-p}}} \left(T \sqrt{\lambda}\right)^{\frac{2}{5-p}}
\ee
It can be checked that by defining an effective dimensionless coupling
constant $\lambda_{\rm eff} = \lambda T^{p-3}$ at temperature $T$, as given 
in \cite{Liu:2006he}, the above expression \eqref{jetq1} can be recast 
precisely into the form given there as,
\be
\hat q = \frac{8 \sqrt{\pi} \Gamma\left(\frac{6-p}{7-p}\right)}
{\Gamma\left(\frac{5-p}{14-2p}\right)}
b_p^{\frac{1}{2}}\lambda^{\frac{p-3}{2(5-p)}}_{\rm eff}(T) \sqrt{\lambda_{\rm
    eff}(T)} \, T^3 \equiv \sqrt{a(\lambda_{\rm eff})}\, 
\sqrt{\lambda_{\rm eff}}
T^3
\ee
where $b_p^{(5-p)/2} =
[2^{16-3p}\pi^{(13-3p)/2}\Gamma((7-p)/2)]/[(7-p)^{7-p}]$ and $a(\lambda_{\rm
  eff})$ characterizes the number of degrees of freedom at temperature $T$.

\section{Conclusion}

To conclude, in this paper using the gravity/gauge theory correspondence and 
the Maldacena prescription we have computed the expectation values of the 
Wilson
loops of $(p+1)$-dimensional strongly coupled Yang-Mills theory. These are
non-perturbative objects and can be related to the observables of quark-gluon
plasma obtained in heavy ion experiments. We have considered both the
time-like and the light-like Wilson loops and used the string probe approach to
compute them. From the time-like Wilson loop we obtained quark-antiquark
separation \eqref{lq} and the velocity dependent quark-antiquark potential
\eqref{potential1} when the dipole moved through the plasma with an arbitrary
velocity $v<1$. As it is hard to write an analytic expressions for them in
general we have plotted these functions in Figures 1, 2, 3. We found that
the general nature of these functions are very similar to $p=3$ obtained in 
\cite{Liu:2006nn,Liu:2006he} except for $p=5$. To see how the details vary for
different $p$'s we have plotted the quark-antiquark separation and
the potential for various values of $p$ at fixed rapidity in 
Figure 4. We have also obtained the form of screening lengths and their 
velocity dependence in \eqref{screenlength}. Although the screening lengths for
general $p$ have been given in \cite{Caceres:2006ta} in the leading order in
rapidity or velocity, we have given the next to leading order corrections to
them. By taking $v \to 1$
limit, the time-like Wilson loop reduces to the light-like Wilson loop and from
there we obtained the jet quenching parameter for the strongly coupled
quark-gluon plasma of $(p+1)$-dimensional Yang-Mills theory whose form was
given earlier in \cite{Liu:2006ug,Liu:2006he}.

\section*{Acknowledgements}

We would like to thank Najmul Haque for helping with 
the numerical integration and plotting the functions in the text.

\end{document}